\documentclass[conference]{IEEEtran}
\IEEEoverridecommandlockouts

\usepackage{cite}
\usepackage{amsmath,amssymb,amsfonts}
\usepackage{algorithm}
\usepackage{algorithmic}
\usepackage{graphicx}
\usepackage{textcomp}
\usepackage{xcolor}

\usepackage{amsthm}
\usepackage{multirow} 
\usepackage{booktabs}
\usepackage{pgfplots}
\usepackage{subfigure}
\usepackage{pifont}
\usepackage{url}

\def\BibTeX{{\rm B\kern-.05em{\sc i\kern-.025em b}\kern-.08em
    T\kern-.1667em\lower.7ex\hbox{E}\kern-.125emX}}

\usepackage{scrextend}
 
\makeatletter
\renewcommand{\footnoterule}{%
  \kern-3pt
  \hrule\@width 1in 
  \@height 0.4pt 
  \kern 2.6pt
}
\makeatother

\begin{document}

\title{Joint Automatic Speech Recognition And Structure Learning For Better Speech Understanding
}


\author{\IEEEauthorblockN{Jiliang Hu\textsuperscript{1}, Zuchao Li\textsuperscript{2,*}, Mengjia Shen\textsuperscript{3}, Haojun Ai\textsuperscript{1}, Sheng Li\textsuperscript{4}, Jun Zhang\textsuperscript{3}}
\IEEEauthorblockA{\textsuperscript{1}Key Laboratory of Aerospace Information Security and Trusted Computing, Ministry of Education, \\School of Cyber Science and Engineering, Wuhan University, Wuhan, China, \\
\textsuperscript{2}School of Computer Science, Wuhan University, Wuhan, China, \\
\textsuperscript{3}Wuhan Second Ship Design and Research Institute, Wuhan, China,\\
\textsuperscript{4}National Institute of Information and Communications Technology, Japan.}


}

\maketitle

\begin{abstract}
Spoken language understanding (SLU) is a structure prediction task in the field of speech. Recently, many works on SLU that treat it as a sequence-to-sequence task have achieved great success. However, This method is not suitable for simultaneous speech recognition and understanding. In this paper, we propose a joint speech recognition and structure learning framework (JSRSL), an end-to-end SLU model based on span, which can accurately transcribe speech and extract structured content simultaneously. We conduct experiments on name entity recognition and intent classification using the Chinese dataset AISHELL-NER and the English dataset SLURP. The results show that our proposed method not only outperforms the traditional sequence-to-sequence method in both transcription and extraction capabilities but also achieves state-of-the-art performance on the two datasets.
\end{abstract}


\renewcommand{\thefootnote}{\fnsymbol{footnote}}
\footnotetext[1]{Corresponding author.}
\renewcommand{\thefootnote}{} 
\footnotetext{This work was supported by the National Natural Science Foundation of China (No. 62306216), the Natural Science Foundation of Hubei Province of China (No. 2023AFB816), the Fundamental Research Funds for the Central Universities (No. 2042023kf0133).}
\footnotetext{Codes are available at \url{https://github.com/193746/JSRSL}}

\section{Introduction}


Automatic speech recognition (ASR) aims to convert human speech into text in the corresponding language \cite{li2022recent}. On the other hand, SLU seeks to enhance machines' ability to comprehend and react to human language by extracting specific structured information from speech. SLU tasks encompass name entity recognition (NER), intent classification (IC), sentiment analysis (SA), among others \cite{arora2024universlu}. SLU models can be categorized into two types: the pipeline model, which uses an ASR module first, followed by a natural language understanding (NLU) module in a sequential manner, and the end-to-end (E2E) model, which directly converts speech representations into structured information. The pipeline model encounters challenges related to error propagation due to the weak linkage between the two modules. Currently, there is a greater emphasis on the E2E model \cite{25}.


There has been significant progress in the field of SLU. However, few works consider how to improve or maintain the accuracy of transcription during spoken language understanding. For NER, most papers utilize the sequence-to-sequence method to achieve entity recognition. For instance, \cite{9} use "\emph{| ]}", "\emph{\$ ]}", and "\emph{\{ ]}" to annotate \emph{Person}, \emph{Place}, and \emph{Organization} in the transcribed text. Subsequently, the annotated text was used to train ASR model. However, \cite{3} have shown that this method of labeling in the transcribed text can affect the recognition performance of the ASR model and increase transcription errors. For classification tasks such as SA and IC, some studies \cite{6,19} also achieve success by annotating classification types in the transcribed text, while others \cite{20,21} directly connect the output of the ASR model to a classification layer. The former will also reduce the performance of transcription, while the latter cannot transcribe the speech simultaneously. It is a significant challenge at the SLU to ensure accurate transcription by the ASR module and correct extraction by the SLU module simultaneously.

In this paper, we propose a joint speech recognition and structure learning framework (JSRSL), an E2E SLU model based on span, which can accurately recognize speech while extracting structured information. This model utilizes a refinement module to enhance text representations for structure prediction. Specifically, we adopt a parallel transformer for non-autoregressive end-to-end speech recognition as our framework and introduce the hidden layer output of the ASR decoder into a span for structure prediction. We additionally introduce a refinement module between the ASR model and the span to strengthen extraction performance. We conduct experiments on NER and IC using the AISHELL-NER and SLURP. The results indicate that the proposed idea has superior capabilities in speech transcription and speech understanding compared to the traditional sequence-to-sequence method. It also outperforms the current state-of-the-art (SOTA) on the two datasets. 



\section{Related Work}

So far, NLU has been well developed, and many works have involved key tasks of natural language processing (NLP), such as natural language inference (NLI) \cite{zhang2020semantics}, semantic role labeling (SRL) \cite{he2018syntax, li2019dependency}, and SA \cite{jing2021seeking,peng2021sparse}. In recent years, a number of excellent SLU systems have emerged. \cite{34} propose an acoustic model called TDT and they attempt to use the proposed model to do SLU tasks by jointly optimizing token prediction and temporal prediction. \cite{20} introduce a novel method known as context-aware fine-tuning. They incorporate a context module into the pre-trained model to extract the context embedding vector, which is subsequently used as extra features for the ultimate prediction. \cite{22} combine pre-trained self-supervised learning (SSL), ASR, language model (LM) and SLU models to explore the model combination that can achieve the best SLU performance and shows that pre-training approaches rooted in self-supervised learning are more potent than those grounded in supervised learning. \cite{4} develop compositional end-to-end SLU systems that initially transform spoken utterances into a series of token representations, followed by token-level classification using a NLU system. \cite{6} suggest a comprehensive approach that combines a multilingual pretrained speech model, a text model, and an adapter to enhance speech understanding. \cite{5} incorporate semantics into speech encoders and present a task-agnostic unsupervised technique for integrating semantic information from large language models (LLMs) \cite{brown2020language,li2023batgpt} into self-supervised speech encoders without the need for labeled audio transcriptions.


\section{Method}
\begin{figure}[htbp]
	\centering 
	\includegraphics[scale=0.7]{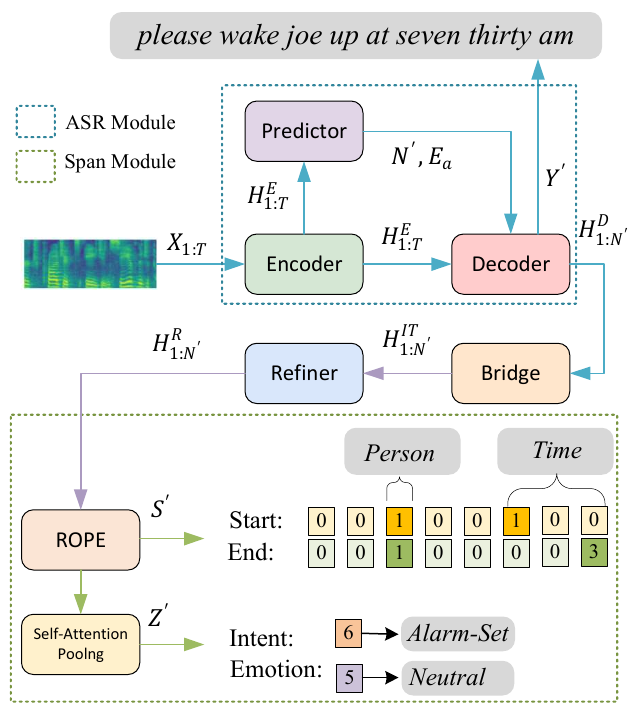} 
	\caption{The structure of our proposed framework, JSRSL.} 
    \label{JSRSL}
\end{figure}

Figure \ref{JSRSL} shows the full overview of our proposed framework, JSRSL. Follow \cite{26}, we adopt this parallel Transformer for non-autoregressive end-to-end speech recognition as our ASR module framework. In addition, we use a \emph{Refiner} module to connect the ASR module and span \cite{2}, and a \emph{Bridge} module to connect the ASR module and \emph{Refiner} module. The \emph{Bridge} module can convert acoustic representation into text representation. The \emph{Refiner} module can strengthen the text representation of upstream inputs and extract richer features for structured learning. Before doing structure learning, we apply rotary embedded encoding (ROPE) \cite{28} into span. ROPE can leverage the boundary information of span by injecting relative position information.

\subsection{SLU Representation Learning}
Let $X$ be a speech sequence with $T$ frames, $X=\{x_1,x_2,x_3,…,x_T\}$. $Y$ is a sequence of tokens, and its length is $N$. Each token is in the vocabulary $V$, $Y=\{y_1,y_2,y_3,…,y_N \mid y_i\in V\}$. The ASR module gets $X$ as input and outputs the decoding result $Y^{'}$ and the \emph{Decoder}'s hidden representation $H_{1:N^{'}}^D$.


The \emph{Refiner} module is essentially a NLU model, and it requires a text vector as input. It will not function correctly if the acoustic hidden representation $H_{1:N^{'}}^D$ is fed directly. If we use the decoded token sequence as input, the computation graph will be truncated, preventing the joint training of the ASR module and \emph{Refiner} module. Therefore, a \emph{Bridge} module is needed to implement the conversion from acoustic hidden representation $H_{1:N^{'}}^D$ to initial text hidden representation $H_{1:N^{'}}^{IT}$. Our \emph{Bridge} module first calculates the probability distribution of the token sequence $Y^{p}$ based on $H_{1:N^{'}}^D$, then it maps the distribution to the embedding layer of the \emph{Refiner} module to convert $Y^{p}$ into the initial text representation $H_{1:N^{'}}^{IT}$. The specific operation of the map is to multiply $Y^{p}$ and the weights of embedding layer $W_{1:V}^{e}$ in a matrix. 

$$
Y^{p}=Softmax(W_{V}^pH_{1:N^{'}}^D+b_{V}^p)
$$
$$
H_{1:N^{'}}^{IT}=Matmul(Y^{p},W_{1:V}^{e})
$$


Subsequently, we feed $H_{1:N^{'}}^{IT}$ into the \emph{Refiner} module for strengthening text representation, resulting in a refined text features $H_{1:N^{'}}^{R}$. It is used as the input features for SLU downstream tasks. The \emph{Refiner} is implemented of multiple stacked Transformer layers.


\subsection{Span-based Structure Learning}
We utilize span to extract the token sequence of each specific information in a given speech sequence $X$ along with their corresponding types. Let $S$ be the pointer sequence, the set of pointer labels be $\Omega$, and $S=\{s_1,s_2,s_3,…,s_N \mid s_i\in \Omega\}$. The pointer sequence corresponds one-to-one to the token sequence obtained from the speech transcription. We define two pointer sequences, namely the starting pointer sequence $S_{start}$ and the ending pointer sequence $S_{end}$. The former is used to locate the head position of the specific information, $\Omega_{start}=\{0,1\}$, and the latter is used to locate the tail position of the specific information and identify its type, $\Omega_{end}=\{0,1,2…N_{type}\}$.





For ordinary structure extraction task, we firstly use a linear layer called $Dense$ to double the dimension of $H_{1:N^{'}}^{R}$, which is aimed at make two pointers use different feature vectors. The output of $Dense$ is evenly divided into two parts, $d_1$ and $d_2$. Then, ROPE is employed to embed relative position information for the pointer sequence.
$$
d_1,d_2=Dense(H_{1:N^{'}}^{R})
$$
$$
S_{start}^{'},S_{end}^{'}=ROPE(d_1),ROPE(d_2)
$$

For classification tasks, we treat them as sentence-level span. Let $Z$ be the classification result, and the set of classification result be $\Omega_{cls}$, $\Omega_{cls}=\{1,2,…,N_{type}\}$. Firstly, we use ROPE to embed relative position information and get the token-level classification features.
$Z_{1:N^{'}}$. 
$$
Z_{1:N^{'}}=ROPE(H_{1:N^{'}}^{R})
$$

Drawing inspiration from \cite{36}, a modified self-attention pooling mechanism is utilized to comprehensively extract sentence-level classification result $Z^{'}$ from token-level features. The process of the self-attention pooling mechanism is as follows, $A_{1:N^{'}}^Z$ is attention value of $Z_{1:N^{'}}$ calculated by a linear layer. 

$$
A_{1:N^{'}}^Z=Softmax(W_{1}Z_{1:N^{'}}+b_{1})
$$
$$
Z^{'}=\underset{N_{type}}{\arg\max}(W_{N_{type}}(\sum_{N^{'}} A_{1:N^{'}}^Z \times Z_{1:N^{'}})+b_{N_{type}})
$$

\subsection{Joint Optimization}
The adopted ASR module calculates three loss functions when training: the cross-entropy (CE), the mean absolute error (MAE), and the minimum word error rate (MWER) loss. CE and MWER are used to optimize the model's transcription ability, while MAE guides the predictor to convergence. According to \cite{26}, the loss function of the ASR part is:
$$
\mathcal{L}_{asr}=\gamma\mathcal{L}_{CE}+\mathcal{L}_{MAE}+\mathcal{L}_{werr}^N(x,y^{*})
$$
$$
\mathcal{L}_{werr}^N(x,y^{*})=\sum_{y_i \in Sample} \widehat{P}(y_i \mid x)[\mathcal{W}(y_i,y^{*})-\widehat{W}]
$$

We also use CE to optimize the model's ability for structure prediction. The total loss function can be formulated as follows, where $\alpha$ is used to control the proportion of ASR loss and structured loss, $\alpha \in (0,1)$.
$$
\mathcal{L}_{total}=(1-\alpha)\mathcal{L}_{asr}+\alpha\mathcal{L}_{sp}
$$

\section{Experiment}
\subsection{Configuration}

We employ AISHELL-NER \cite{3} and SLURP's NER (slot filling) subset for NER, and SLURP's IC subset for IC \cite{29}. Following the previous work \cite{23} that used SLURP, the provided synthetic data is utilized for training. For assessment metrics, SLURP-NER utilizes WER and SLURP-F1, SLURP-IC employs WER and micro-F1, while AISHELL-NER uses CER and micro-F1. Our baseline, Seq2Seq, applies the sequence-to-sequence method to the JSRSL's ASR module and annotates like \cite{3}. Another baseline, Pipe, is the concatenation of the JSRSL's ASR module and a NLU module, predicting structure by span. The models chosen as benchmarks all adopt sequence generation method by adding special symbols in transcribed text or generating structured sequence directly.

We build the experiment environment based on Funasr \cite{30} and ModelScope, utilizing 220M Chinese and English Paraformer for experiments. The pre-training parameters of BERT-base \cite{31} is adopted to initialize the \emph{Refiner} of JSRSL and the NLU module of Pipe. The $\alpha$ of the loss is set to 0.5. The models are trained until converges. We consistently use the Adam optimizer with a learning rate of 5e-5. 


	

\subsection{Main Result}

\begin{table}[htbp] 
    \renewcommand{\arraystretch}{1.2} 
    \small 
    \caption{Comparison results with benchmarks in AISHELL-NER and SLURP. In SLURP, all models conduct joint training and evaluation for NER and IC tasks. The BERT in \cite{3}'s pipeline model are pre-trained. The Post AED uses 1k external data.}
    \label{main results}
	\resizebox{\linewidth}{!}{
		\begin{tabular}{ccccc}	 
			\hline
			\multirow{2}{*}[-1.5ex]{Model} &\multicolumn{2}{c}{\multirow{2}{*}[-1.5ex]{Paradigm}}&\multicolumn{2}{c}{\multirow{1}{*}{AISHELL-NER}} \\ 
			\cmidrule(lr){4-5}
			&&& CER ($\downarrow$) & F1 ($\uparrow$)\\
			\hline
			
			Transformer \cite{3} &\multicolumn{2}{c}{E2E}&9.25&64.34 \\
			Conformer \cite{3} &\multicolumn{2}{c}{E2E}&4.79&73.37 \\
			Transformer+BERT* \cite{3} &\multicolumn{2}{c}{Pipeline}&9.25&65.95\\
			Conformer+BERT* \cite{3} &\multicolumn{2}{c}{Pipeline}&4.79&74.90\\	
			
			\hline
			Seq2Seq (Ours) &\multicolumn{2}{c}{E2E}&3.48&76.17 \\
            Pipe (Ours)  &\multicolumn{2}{c}{Pipeline}&1.76&77.98 \\
			JSRSL (Ours)&\multicolumn{2}{c}{E2E}&\textbf{1.71}&\textbf{80.85} \\ 
			
			\hline
			\multirow{2}{*}[-1.5ex]{Model} &\multirow{2}{*}[-1.5ex]{Paradigm}&\multicolumn{3}{c}{\multirow{1}{*}{SLURP}}  \\
			\cmidrule(lr){3-5}
			&& WER ($\downarrow$) & SLU F1 ($\uparrow$) & IC F1 ($\uparrow$)\\
			\hline
			RNNT$\rightarrow$BertNLU \cite{raju2021joint}&Pipeline &15.20&72.35 &82.45 \\
			Espnet SLU \cite{23} &E2E &-&71.90 &86.30 \\
			PostDec AED* \cite{6} &E2E &-&80.08&89.33\\
			Conformer-RNNT \cite{19}  &E2E &14.20&77.22&90.10\\
			Conformer-CTC \cite{19}  &E2E &14.50&69.30&85.50\\
			TDT 0-6 \cite{34} &E2E &-&\textbf{80.61}&89.28 \\
			TDT 0-8 \cite{34} &E2E &-&79.90&90.07 \\
			
			\hline
			Seq2Seq (Ours)  &E2E &18.83&68.84&86.63 \\
            Pipe (Ours)  &Pipeline&13.61&76.62&86.67 \\
			JSRSL (Ours) &E2E &\textbf{13.14}&\textbf{80.17}&\textbf{91.03}\\
			\hline

		\end{tabular}
	}
	\centering
\end{table}

\begin{table*}[htbp]
    \renewcommand{\arraystretch}{1.1}
	\large
    \caption{Ablation results in AISHELL-NER and SLURP. Sampler is a component in Paraformer used to enhance training effectiveness.}
    \label{ablation results}
	\resizebox{\linewidth}{!}{
		\begin{tabular}{ccccccccccccccc}	 
			\hline
			\multicolumn{4}{l}{\multirow{2}{*}{Model}} & &\multicolumn{4}{c}{AISHELL-NER} &\multicolumn{4}{c}{SLURP-NER} & \multicolumn{2}{c}{SLURP-IC}\\
			\cmidrule(lr){6-9}\cmidrule(lr){10-13}\cmidrule{14-15}
			&&&&Refiner& CER ($\downarrow$) & Precision ($\uparrow$) & Recall ($\uparrow$) & F1 ($\uparrow$) &WER ($\downarrow$) & Precision ($\uparrow$) & Recall ($\uparrow$) & SLU F1 ($\uparrow$) & WER ($\downarrow$) & F1 ($\uparrow$) \\
			\hline
			\multicolumn{4}{l}{JSRSL} &\ding{53}&1.76&\textbf{83.21}&\textbf{78.15}&\textbf{80.46} &10.79&\textbf{80.25}&73.37&\textbf{76.66}  &\textbf{13.84}&\textbf{88.19}\\ 
			\multicolumn{4}{l}{\quad\quad\quad w/o ROPE} &\ding{53}&1.78&82.50&78.00&80.06 &10.81&79.33&\textbf{73.44}&76.27  &13.87&87.93\\ 
			
			\multicolumn{4}{l}{\quad\quad\quad w/o Sampler} &\ding{53}&\textbf{1.75}&82.98&78.00&80.32  &\textbf{10.72}&80.14&73.20&76.51 &14.05&\textbf{88.19}\\ 
			\hline
			\multicolumn{4}{l}{JSRSL} &\ding{51}&1.77&81.89&79.27&80.45 &\textbf{11.17}&\textbf{83.26}&\textbf{77.46}&\textbf{80.26}  &\textbf{13.30}&\textbf{89.90}\\ 
			\multicolumn{4}{l}{\quad\quad\quad w/o pretrain} &\ding{51}&\textbf{1.71}&\textbf{82.36}&\textbf{79.54}&\textbf{80.85} &12.17&82.41&76.36&79.27  &13.53&89.67\\ 
			\hline
			
		\end{tabular}
	}
	\centering
\end{table*}

Table \ref{main results} presents the comparison results with baselines and benchmarks. Before the joint training of Paraformer and BERT, the Paraformer in JSRSL is pre-trained by audio-text pairs from AISHELL-NER or SLURP and the BERT is pre-trained by the vocabulary of Paraformer and the text-span pairs of corresponding dataset. The results show that the proposed model, based on span, outperforms the E2E and pipeline baselines and benchmarks using the sequence-to-sequence method in both structure extraction and ASR. By using span, we can distinguish between ASR and SLU tasks, avoid the impact of extra structured annotations on transcription accuracy and allow for more precise extraction of structured information. In addition, Pipe is worse than JSRSL due to issues with error propagation of pipeline model, indicating that the E2E model is more promising. In SLURP, TDT models and PostDEC AED have the best SLU performance in the benchmarks. TDT models adopt the token and duration transformer architecture, which enhances the ability of sequence generation and accelerates the inference speed of the original RNN-Transducer \cite{ghodsi2020rnn} through joint learning of token and duration prediction. "0-X" refers to the maximum duration of duration configurations as "X". The performance of TDT is comparable to that of our JSRSL, but it is influenced by specific configurations. TDT 0-6 performs better on SLURP-NER, while TDT 0-8 performs better on SLURP-IC. PostDec AED uses an adaptor to concatenate the speech encoder and NLU model. The adaptor tries to minimize the distance between the speech representation and the text representation. Before finetune the SLU system, it must undergo adapter pretraining, which optimizes the loss between predicted text and real text. Although PostDec AED's performance is just slightly inferior to our JSRSL, its training process is more complex and requires a large amount of external data.





\subsection{Ablation Study}

\begin{table}[htbp] 
	\small
    \renewcommand{\arraystretch}{1.1}
    \caption{ASR results (CER on AISHELL-NER and WER on SLURP) among Paraformer, Whisper, XLSR-53 and our JSRSL with similar size.}
    \label{asr results}
		\begin{tabular}{cccc} 
			\hline
			Model&Params&AISHELL-NER&SLURP \\
			\hline
            XLSR-53 \cite{babu2021xls}&315M&4.12&22.10\\ 
			Whisper \cite{32}&244M&5.54&16.89 \\
            Paraformer&220M&\textbf{1.76}&\textbf{13.61} \\
			\hline
            JSRSL (Ours) &314M&\textbf{1.71}&\textbf{13.14} \\ 
            \hline
		\end{tabular}
	\centering
\end{table}

Table \ref{ablation results} shows the ablation results. The proposed model shows a certain improvement after introducing ROPE. In addition, in SLURP-NER and AISHELL-NER, we find that \emph{Sampler} does not significantly improve model's ASR performance, but rather increases its spoken understanding ability. In AISHELL-NER and SLURP-IC, the introduction of refinement module leads to a further decrease in the model's CER or WER and an increase in F1 score. However, in SLURP-NER, we find that with refinement module, while JSRSL enhances SLU F1, WER increases. But overall, the introduction of refinement modules is beneficial for improving model performance. In AISHELL-NER, the JSRSL with pre-trained Paraformer does not achieve better performance, possibly due to overfitting. Nevertheless, in SLURP, pre-trained ASR module improves the overall ability of JSRSL. Additionally, to assess the effectiveness of the adopted ASR module, we carry out an ASR experiment detailed in Table \ref{asr results}. The results indicate that Paraformer's ASR performance on AISHELL-NER and SLURP surpasses that of Whisper and XLSR-53. This demonstrates the robustness of our model's ASR module. Furthermore, JSRSL achieves better ASR capability after joint training.

\subsection{OOD Analysis}

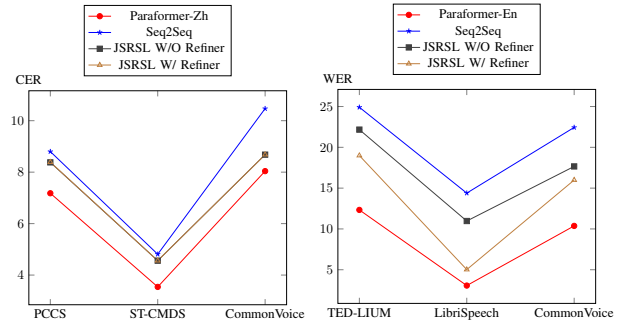
\begin{figure}[htbp]
	\centering 	
	\begin{minipage}{0.45\linewidth}
		\centering
		\begin{tikzpicture}[scale=0.5]	
			\begin{axis}[
				sharp plot,
				legend style={at={(0.5,1.4)},anchor=north},
				every axis y label/.style={at={(0,1)}, above=0.5mm},
				symbolic x coords={PCCS,ST-CMDS,CommonVoice},
				xtick=data,
				ylabel={CER}] 
				
				\addplot[color=red,mark=*] coordinates{
					(PCCS,7.18)
					(ST-CMDS,3.54)
					(CommonVoice,8.04)
				};
				\addlegendentry{Paraformer-Zh}
				
				\addplot[color=blue,mark=star] coordinates{
					(PCCS,8.80)
					(ST-CMDS,4.81)
					(CommonVoice,10.47)
				};
				\addlegendentry{Seq2Seq}
				
				\addplot[color=darkgray,mark=square*] coordinates{
					(PCCS,8.38)
					(ST-CMDS,4.56)
					(CommonVoice,8.68)
				};
				\addlegendentry{JSRSL W/O Refiner}
				
				\addplot[color=brown,mark=triangle] coordinates{
					(PCCS,8.39)
					(ST-CMDS,4.56)
					(CommonVoice,8.68)
				};
				\addlegendentry{JSRSL W/ Refiner}
			\end{axis}
		\end{tikzpicture}
	\end{minipage}
	\begin{minipage}{0.45\linewidth}
		\centering
		\begin{tikzpicture}[scale=0.5]	
			\begin{axis}[
				sharp plot,
				legend style={at={(0.5,1.4)},anchor=north},
				every axis y label/.style={at={(0,1)}, above=0.5mm},
				symbolic x coords={TED-LIUM,LibriSpeech,CommonVoice},
				xtick=data,
				ylabel={WER}] 
				
				\addplot[color=red,mark=*] coordinates{
					(TED-LIUM,12.33)
					(LibriSpeech,3.06)
					(CommonVoice,10.37)
				};
				\addlegendentry{Paraformer-En}
				
				\addplot[color=blue,mark=star] coordinates{
					(TED-LIUM,24.91)
					(LibriSpeech,14.39)
					(CommonVoice,22.44)
				};
				\addlegendentry{Seq2Seq}
				
				\addplot[color=darkgray,mark=square*] coordinates{
					(TED-LIUM,22.17)
					(LibriSpeech,10.97)
					(CommonVoice,17.66)
				};
				\addlegendentry{JSRSL W/O Refiner}
				
				\addplot[color=brown,mark=triangle] coordinates{
					(TED-LIUM,18.99)
					(LibriSpeech,5.02)
					(CommonVoice,15.99)
				};
				\addlegendentry{JSRSL W/ Refiner}
			\end{axis}
		\end{tikzpicture}
	\end{minipage}
	
	\caption{Out-of-Distribution experiment result. The results on the left are in the Chinese datasets and the results on the right are in the English datasets.}	
    \label{ood results}
\end{figure}

			

This section conducts Out-of-Distribution (OOD) experiment with unseen ASR datasets. The Chinese ASR datasets are Primewords Chinese Corpus Set 1 (PCCS) \cite{primewords_201801}, Free ST Chinese Mandarin Corpus (ST-CMDS), and Common Voice (Zh) \cite{commonvoice:2020}. The English ASR datasets include Librispeech \cite{panayotov2015librispeech}, Ted-LIUM \cite{rousseau2012ted}, and Common Voice (En). Each dataset uniformly uses the first 8000 training samples for the experiment. When evaluating the Seq2Seq, the special symbols that indicate name entities are ignored. Figure \ref{ood results} presents the OOD results. The original Chinese or English Paraformer demonstrates better generalization ability and achieves the lowest CER or WER among all datasets. Although all trained models' transcription performance has decreased, compared to Seq2Seq, which uses a sequence generation method, the span-based models exhibit lower recognition error rates. What's more, JSRSL with refinement module performs better than that without refinement module. This indicates that the proposed method results in less performance loss in speech recognition, making it more suitable for joint ASR and structure prediction.

\section{Conclusion}
This paper proposes a method called JSRSL for jointly ASR and structure prediction based on span. This approach aims to ensure accurate transcribing and understanding of speech simultaneously. Experiments have shown that the proposed scheme is superior to traditional sequence-to-sequence method in both transcription and extraction capabilities, and achieves SOTA performance on the datasets used for the experiments.

\clearpage
\bibliographystyle{IEEEtran}
\bibliography{custom}

\end{document}